\documentclass{sistedes}
\usepackage[english]{babel}
\usepackage{graphicx}
\usepackage{subcaption}
\begin{document}
\title{AI-Driven Multi-Region Provisioning for Cloud Services Using Spot Fleets}
%
%
\author{Javier Fabra\inst{1}\orcidID{0000-0001-5549-7649} \and
Enrique Molina-Gim\'enez\inst{2}\orcidID{0009-0005-2597-3815} \and
Pedro Garc\'ia-L\'opez\inst{2}\orcidID{0000-0002-9848-1492}}
\authorrunning{J. Fabra et al.}
%
\institute{
Aragon Institute for Engineering Research (I3A),\\
Department of Computer Science and Systems Engineering,\\
Universidad de Zaragoza, Zaragoza, Spain\\
\email{jfabra@unizar.es}\\
\and
Departament d'Enginyeria Informàtica, Matemàtiques\\ 
Universitat Rovira i Virgili, Tarragona, Spain \\
\email{\{enrique.molina,pedro.garcia\}@urv.cat}
}

\maketitle              
\begin{abstract}
Cloud service platforms increasingly rely on elastic infrastructures to support dynamic workloads. Spot instances provide discounted computing resources but introduce uncertainty due to dynamic pricing, resource availability, and interruption risks that vary across geographical regions. In Amazon Web Services, the EC2 Spot Service simplifies fleet provisioning through allocation strategies, but it cannot estimate fleet costs before deployment and restricts provisioning to a single region. This paper presents an AI-driven provisioning service for multi-region spot fleets. The proposed approach combines monitoring of provisioning plans with predictive models to estimate fleet configurations and prices before launch, enabling cost-aware deployment decisions across regions while preserving the operational behavior of the EC2 Spot Service. The system was validated with fleets of up to 1500 vCPUs. Experimental results show a prediction accuracy of 99.79\% compared to the EC2 Spot Service and potential cost savings of up to 64\% by exploiting regional price variability.

\end{abstract}

\section{Introduction}\label{sec:intro}

Cloud providers offer discounted virtual machines when unused capacity is available. These resources, commonly known as \textit{spot instances}, provide significantly lower prices than on-demand instances, but introduce uncertainty due to dynamic pricing, fluctuating availability, and possible interruptions. Despite these limitations, spot instances are widely used in workloads that tolerate failures or short-lived interruptions, such as batch processing, data analytics, and large-scale parallel applications \cite{seespot,deepspotcloud,flint}.

In Amazon Web Services (AWS), spot fleets can be provisioned through the EC2 Spot Service, which offers several allocation strategies to balance cost and reliability. However, this service presents two important limitations for service-oriented cloud deployments: \textit{(i)} fleet cost cannot be estimated before launch, and \textit{(ii)} provisioning is restricted to a single region, preventing users from exploiting regional price differences. These limitations hinder cost-aware orchestration of cloud services, especially for applications requiring dynamic scaling or large computational resources.

From a service engineering perspective, provisioning decisions play a central role in the operational behavior of cloud services. Many modern service platforms rely on elastic infrastructures to dynamically allocate compute resources depending on workload demand. In such scenarios, provisioning strategies directly influence not only infrastructure cost but also service reliability and scalability.

However, deciding where and how to provision spot fleets is a complex task. Cloud markets exhibit strong regional variability in pricing and capacity, while allocation strategies introduce trade-offs between cost efficiency and interruption risk. As a result, service operators often lack sufficient information to determine the most advantageous deployment region before launching their workloads. This challenge motivates the need for intelligent provisioning mechanisms capable of estimating deployment alternatives in advance and guiding cost-aware orchestration decisions.

This paper presents an AI-driven provisioning service that supports multi-region provisioning of spot fleets. The approach combines monitoring of AWS provisioning behavior with predictive models that estimate fleet configurations and prices before deployment. In this way, the system supports region-aware, cost-conscious provisioning decisions while preserving the provisioning logic of the EC2 Spot Service.

The main contributions of this work are:
\begin{enumerate}
    \item An AI-driven service for estimating provisioning plans and costs of heterogeneous spot fleets before deployment.
    \item A multi-region provisioning approach that enables cost-aware deployment decisions for cloud services.
    \item An experimental evaluation on AWS showing 99.79\% prediction accuracy with respect to the EC2 Spot Service and savings of up to 64\% across regions.
\end{enumerate}

The rest of the paper is organized as follows. Section~\ref{sec:context} presents the problem context and related work. Section~\ref{sec:approach} describes the proposed service architecture and implementation. Section~\ref{sec:eval} evaluates the approach. Finally, Section~\ref{sec:conclusions} concludes the paper.

\section{Problem context and related work}\label{sec:context}

Spot instances are offered by major cloud providers as discounted resources subject to possible interruption. In AWS, spot provisioning is performed through pools defined by instance type and availability zone, where pricing and allocation depend on demand and real-time capacity. To simplify provisioning, AWS provides strategies such as \textit{capacity-optimized}, \textit{lowest-price}, and \textit{price-capacity-optimized}. While these strategies are useful, they are restricted to a single region and provide no estimation of total fleet cost before deployment.

This variability is particularly relevant for applications that deploy large compute fleets. In these scenarios, even small differences in instance pricing or capacity availability may lead to substantial cost variations at the fleet level. Furthermore, provisioning decisions are typically taken under incomplete information, since cloud providers do not expose mechanisms to estimate the resulting fleet composition before the request is executed. Consequently, users often rely on trial-and-error provisioning or static regional choices that do not fully exploit the economic dynamics of the cloud market.

\begin{figure}[htbp!]
  \centering
  \begin{minipage}{0.48\textwidth}
      \centering
      \includegraphics[width=\linewidth]{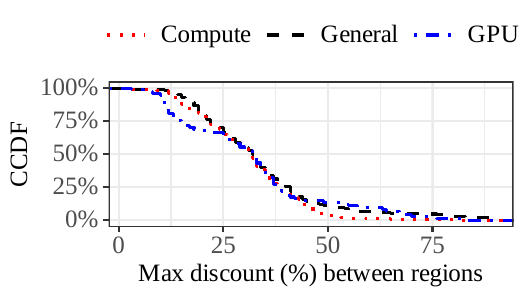}
      \caption{Maximum price difference between AWS availability zones.}
      \label{fig:discounts}
  \end{minipage}
  \hfill
  \begin{minipage}{0.48\textwidth}
      \centering
      \includegraphics[width=\linewidth]{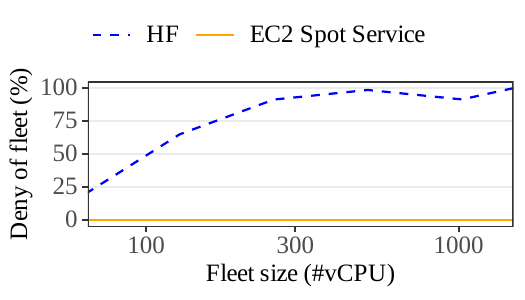}
      \caption{Denial rate of homogeneous fleets in AWS.}
      \label{fig:homo}
  \end{minipage}
\end{figure}

A central motivation for a multi-region provisioning service is the high variability of spot prices across geographical locations. Spot prices fluctuate not only over time but also across regions and availability zones, creating opportunities for substantial savings when workloads can be placed in the most advantageous region. Figure~\ref{fig:discounts} shows that the maximum variation between the cheapest and most expensive locations is substantial for a relevant subset of instance types. In addition, naive provisioning based on a single cheap instance type is not robust at fleet scale. As shown in Figure~\ref{fig:homo}, homogeneous fleet construction leads to a high denial rate as target capacity increases, motivating the need for predictive and heterogeneous provisioning.

Previous research has addressed complementary aspects of spot computing. Some works focus on spot price prediction and cost reduction through statistical and machine learning techniques \cite{fabra-spots-fgcs,predspot,convolution_spot}. Others study availability and interruption behavior \cite{www-kylee,spotlake}. Spot instances have also been integrated into higher-level distributed systems such as Hadoop, Spark, and GPU-based training platforms \cite{seespot,deepspotcloud,flint,autobot}. Finally, several approaches exploit cross-region or multi-cloud resource variability, such as DeepSpotCloud, SpotVerse, SkyPilot, and SkyServe \cite{deepspotcloud,spotverse,skypilot,skyserve}.

However, to the best of our knowledge, existing approaches do not jointly address the problem of predicting robust provisioning plans for large heterogeneous spot fleets before deployment and across multiple regions. This gap is especially relevant in the context of cloud services, where provisioning decisions directly affect cost, scalability, and operational quality.

\section{AI-driven provisioning service}\label{sec:approach}

\begin{figure}[htbp!]
    \centering
    \includegraphics[width=\linewidth]{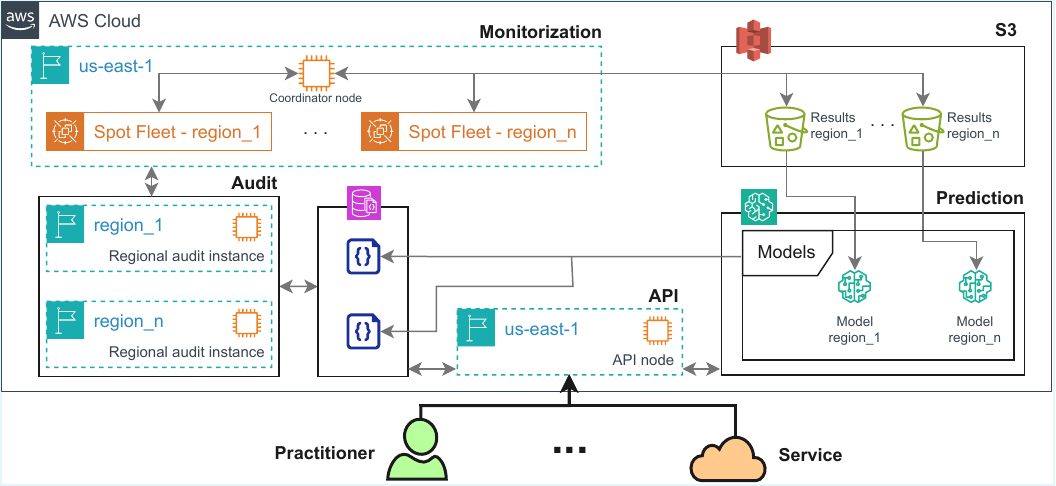}
    \caption{Deployment overview of the proposed AI-driven provisioning service on AWS.}
    \label{fig:deploy}
\end{figure}

The proposed solution is designed as a service-oriented architecture for predictive multi-region provisioning. Its goal is to estimate, before launch, the fleet configuration and expected cost associated with a given target capacity and allocation strategy. The architecture is organized around four logical components: monitoring, prediction, auditing, and API access. Figure~\ref{fig:deploy} presents an overview of the deployment of these components on AWS.

From a service architecture perspective, the system can be understood as an intelligent provisioning layer that operates on top of existing cloud APIs. Instead of replacing the underlying provisioning mechanism, the proposed service augments it with predictive capabilities that enable proactive decision making. This design choice allows the framework to remain compatible with existing cloud infrastructures while providing additional intelligence to guide deployment strategies.

Another key design principle is modularity. Each component of the architecture addresses a specific aspect of the provisioning process: monitoring collects empirical observations of cloud behavior, prediction models learn provisioning patterns, the API exposes provisioning recommendations as a service interface, and the audit module maintains model validity over time. This separation of concerns simplifies system evolution and facilitates integration with higher-level service orchestration platforms.

The \textit{monitoring component} probes the EC2 Spot Service by issuing provisioning requests for different fleet sizes, strategies, and regions. The resulting plans are stored together with metadata such as region, timestamp, instance types, and estimated cost. This dataset captures the temporal and regional behavior of spot provisioning and constitutes the basis for model training.

These observations allow the predictive models to approximate the provisioning decisions normally taken by the EC2 Spot Service, enabling the system to estimate fleet composition and cost before the actual request is issued.

The \textit{prediction component} uses this monitored data to train LSTM-based models that estimate provisioning plans for each region and strategy. Instead of relying on a single global predictor, the system trains one specialized model per region and allocation strategy. This design improves adaptation to local market dynamics and enables more precise cost estimation.

LSTM models were selected because provisioning plans and spot prices exhibit temporal dependencies. Their recurrent structure allows the model to capture time-based patterns in pricing and availability that traditional regression models cannot easily represent.

The \textit{API component} exposes the service to users and applications. Requests specify the target capacity, the desired allocation strategy, and optionally a region. If no region is specified, the system evaluates all supported regions and returns a ranked list of provisioning alternatives, including estimated cost, instance composition, and deployment location. This service-oriented interface facilitates the integration of predictive cloud provisioning into higher-level service platforms and orchestration workflows.

Finally, the \textit{audit component} validates the recommendations by periodically comparing predictions with actual provisioning outcomes. When the deviation exceeds a predefined threshold, the monitoring phase is reactivated and the corresponding models are retrained. This feedback loop keeps the service aligned with the changing behavior of the cloud market.

The current implementation has been deployed on AWS and focuses on modern x86 compute instances from the \texttt{c6i} and \texttt{c7i} families. An initial monitoring campaign launched 720 fleets per region across nine AWS regions, covering target capacities from 64 to 1500 vCPUs during a 90-day period. The collected data was stored in S3, used to train region-specific LSTM models in SageMaker, and served through an API deployed on EC2. Although implemented on AWS, the architecture is sufficiently modular to be extended to other providers.

\section{Evaluation}\label{sec:eval}

The proposed service was evaluated with three objectives: \textit{(i)} verifying that the recommended fleets can be successfully provisioned, \textit{(ii)} measuring the agreement with the EC2 Spot Service, and \textit{(iii)} quantifying the savings enabled by multi-region deployment. After the initial monitoring phase, the evaluation was carried out over seven days, from January 24 to January 30, 2025. During this period, the system launched 10 fleets every three hours, with randomly selected sizes and allocation strategies, for a total of 720 experiments.

All experiments were executed using the same AWS configuration and API interactions used by production spot fleets, ensuring that the evaluation reflects real provisioning behavior.

\subsection{Provisioning success and prediction accuracy}

The first result is that none of the recommended fleets was rejected by AWS. This indicates that the proposed approach preserves the practical feasibility of the EC2 Spot Service while moving provisioning decisions to a predictive and multi-region setting.

To assess accuracy, we compared the recommended provisioning plans with those returned by the EC2 Spot Service under the same request conditions. Table~\ref{table} summarizes the mismatch rate across fleet sizes. In 92.78\% of the cases, the predicted plan exactly matched the EC2 recommendation. When mismatches occurred, they did not affect the fulfillment rate and produced an average price deviation of only 2.97\%. Overall, this corresponds to a prediction accuracy of 99.79\% with respect to the EC2 Spot Service.

\begin{table}[h]
\centering
\begin{tabular}{lrrr}
\hline
Size (\#vCPU) & Mismatch & Total & Error rate \\ \hline
64   & 6  & 120 & 5.00\% \\
128  & 7  & 120 & 5.83\% \\
256  & 11 & 120 & 9.17\% \\
512  & 10 & 120 & 8.33\% \\
1200 & 9  & 120 & 7.50\% \\
1500 & 9  & 120 & 7.50\% \\ \hline
General & 52 & 720 & 7.22\% \\ \hline
\end{tabular}
\caption{Mismatch rate between the predictive model and the EC2 Spot Service.}
\label{table}
\end{table}

\subsection{Multi-region savings}

\begin{figure}[htbp!]
    \centering
    \includegraphics[width=0.8\linewidth]{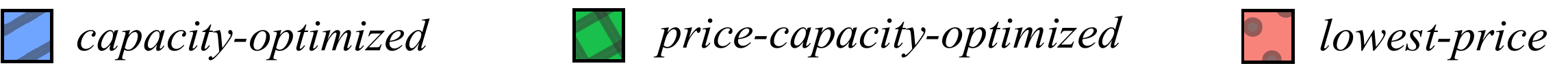}

    \vspace{0.5em}

    \begin{subfigure}[b]{0.32\textwidth}
        \centering
        \includegraphics[width=\linewidth]{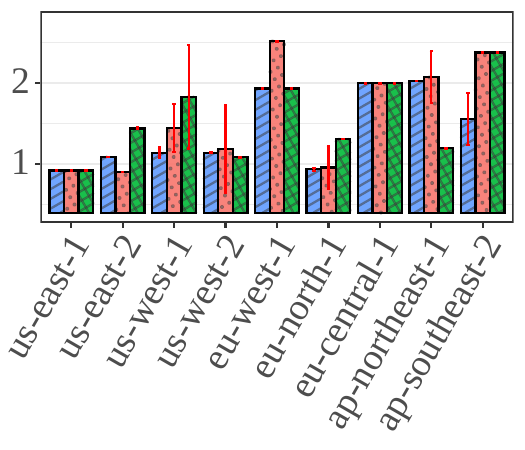}
        \caption{64 vCPU}
        \label{fig:any-region-64}
    \end{subfigure}
    \hfill
    \begin{subfigure}[b]{0.32\textwidth}
        \centering
        \includegraphics[width=\linewidth]{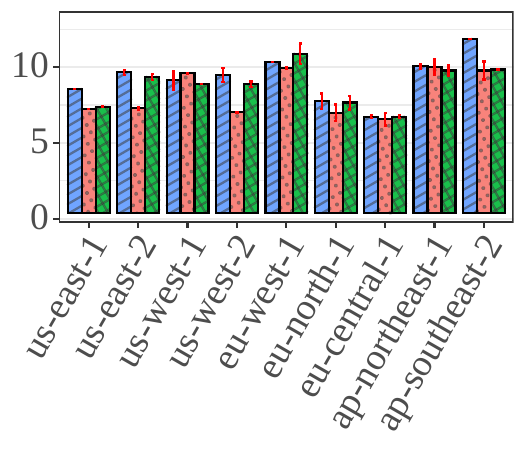}
        \caption{512 vCPU}
        \label{fig:any-region-512}
    \end{subfigure}
    \hfill
    \begin{subfigure}[b]{0.32\textwidth}
        \centering
        \includegraphics[width=\linewidth]{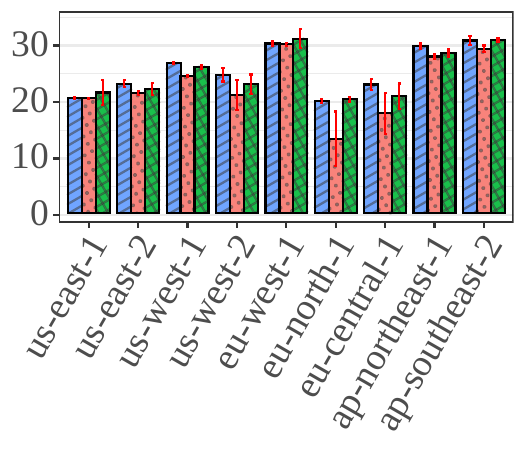}
        \caption{1500 vCPU}
        \label{fig:any-region-1500}
    \end{subfigure}

    \caption{Estimated fleet cost across regions for different target capacities.}
    \label{fig:any-region}
\end{figure}

A key advantage of the service is the ability to estimate fleet costs across regions before deployment. Figure~\ref{fig:any-region} shows the price differences observed for fleets of 64, 512, and 1500 vCPUs across the nine monitored regions. The results confirm that regional price differences remain significant at all scales. For instance, the same 64 vCPU fleet can differ by a factor of 2.77$\times$ between regions, while larger fleets still exhibit substantial savings.

The maximum observed savings reached 64\%, confirming that region-aware provisioning is a major source of cost optimization. Importantly, these savings are considerably larger than the differences introduced by changing the allocation strategy within the same region, which were below 10\% on average in our experiments.

\subsection{Observations for service provisioning}

Beyond prediction accuracy, the experiments reveal three practical insights for cloud service engineering. First, AWS may overprovision fleets when the target capacity is expressed in computational units, especially with the \textit{lowest-price} strategy. In some cases, this behavior makes \textit{lowest-price} more expensive than more conservative strategies. Second, regional price differences remain consistent over time, suggesting that predictive services can exploit stable cost asymmetries across locations. Third, eviction behavior differs sharply across strategies: in our experiments, \textit{lowest-price} approached 94.9\% eviction probability after one hour, while \textit{capacity-optimized} remained substantially more stable. These findings reinforce the need for service-level provisioning mechanisms that jointly consider price, capacity, and regional variability.

From a service management perspective, these results highlight the relevance of predictive provisioning for cost-aware cloud orchestration. In traditional provisioning workflows, deployment decisions are often taken without visibility of the final infrastructure composition. By estimating provisioning outcomes before launch, the proposed approach allows service operators to compare alternative deployment scenarios and select the most advantageous region.

Furthermore, the observed regional price variability suggests that global provisioning strategies can significantly improve resource efficiency in cloud services. Even when allocation strategies within a region produce relatively small price differences, the selection of the deployment region itself may lead to substantially larger savings. These findings reinforce the importance of integrating predictive models into service-level infrastructure management.

\section{Conclusions}\label{sec:conclusions}

This paper has presented an AI-driven provisioning service for multi-region spot fleets in cloud environments. The approach extends the capabilities of the EC2 Spot Service by enabling price estimation before deployment and by supporting region-aware provisioning decisions. The evaluation on AWS shows that the proposed system preserves provisioning feasibility, achieves 99.79\% accuracy with respect to the EC2 Spot Service, and enables savings of up to 64\% by exploiting regional differences.

Beyond the specific case of AWS spot fleets, this work shows how predictive intelligence can be integrated into service-oriented cloud provisioning workflows. By exposing provisioning recommendations through an API interface, the system can support higher-level orchestration layers in complex service ecosystems.

The resulting system can support higher-level service platforms that need to orchestrate compute resources under cost and scalability constraints. As future work, we plan to extend the approach to other instance families and cloud providers, and to study how predictive provisioning can be integrated into ap\-pli\-ca\-tion-level service orchestration.

%
%
%
%

\bibliographystyle{IEEEtran}
\bibliography{references}

\end{document}